\documentclass[preprint, 12pt]{elsarticle}

\makeatletter
\def\ps@pprintTitle{%
  \let\@oddhead\@empty
  \let\@evenhead\@empty
  \let\@oddfoot\@empty
  \let\@evenfoot\@oddfoot
}
\makeatother

\usepackage{graphicx}
\usepackage{adjustbox}
\usepackage{multirow}
\usepackage{hyperref}
\usepackage[table,xcdraw]{xcolor}
\usepackage{rotating}
\usepackage{caption}
\usepackage{amsmath}
\usepackage{amsfonts}
\usepackage{float}

\begin{document}

\begin{frontmatter}

\title{Reimagining Application User Interface (UI) Design using Deep Learning Methods: Challenges and Opportunities}

\author[add1]{Subtain Malik}
\ead{m3sibti@gmail.com}
\author[add1]{Muhammad Tariq Saeed}
\ead{tariq@sines.nust.edu.pk}
\author[add1]{Marya Jabeen Zia}
\ead{mjzia23@gmail.com}
\author[add1]{Shahzad Rasool}
\ead{shahzad.rasool@sines.nust.edu.pk}
\author[add2]{Liaquat Ali Khan}
\ead{liaquatalikhan@gmail.com}
\author[add1]{Mian Ilyas Ahmed}
\ead{m.ilyas@rcms.nust.edu.pk}

\address[add1]{School of Interdisciplinary Engineering and Sciences (SINES), National University of Sciences and Technology (NUST), Islamabad, Pakistan}
\address[add2]{Air University, Islamabad, Pakistan}

\begin{abstract}
In this paper, we present a review of the recent work in deep learning methods for user interface design. The survey encompasses well known deep learning techniques (deep neural networks, convolutional neural networks, recurrent neural networks, autoencoders, and generative adversarial networks) and datasets widely used to design user interface applications. We highlight important problems and emerging research frontiers in this field. We believe that the use of deep learning for user interface design automation tasks could be one of the high potential fields for the advancement of the software development industry. 
\end{abstract}

\begin{keyword}
Deep Learning \sep Human-Computer Interaction \sep User Interface Designs \sep Data-Driven Models
\end{keyword}

\end{frontmatter}


\section{Introduction}
Interactive systems have transitioned from being used in specialized environments to being ingrained into everyday life. All sorts of devices including ticketing machines, home appliances, kiosks, wearable devices, etc. have computing devices  embedded in them. Thus, the design of interactive systems has evolved from creating 'standalone' systems to designing whole environments where the context of system usage is an essential part of the design consideration. A number of factors contribute towards the success of an interactive system which is often evaluated in terms of usability, accessibility, learnability and memorability among others. The user interface (UI) design remains a critical part of the user experience (UX) design. The conventional process of UI design involves understanding aspirations of the user and studying activities in context of the application. It requires exploration and design of technological solutions that fit the users, the activities, and the context of their usage. The process also involves evaluation of design alternatives that are refined in an iterative manner. A range of skills from multiple disciplines are involved, and a single designer seldom carries all the skills needed for the design of such a solution.
\par
The user centered approach to the design of interfaces allows for engineering design alternatives most likely to succeed. Nonetheless, much of the UI design process is still guided by best practices and principles that originate from various disciplines such as psychology, human factors, ergonomics, etc. As a result a number of established design guidelines have come into existence which significantly reduce the time, cost, and effort to arrive at a feasible technological solution and a successful UI design.
\par
The idea of generative UI modelling can be traced back to pattern based models for UI design and code generation approaches. In the past, several contributions have been made by using various statistical and machine learning (ML) techniques for UI design. However, a major limitation of many ML algorithms is the requirement for hard-coded rules. 
For instance, while recognizing a cat or dog, a human infant learns the classes of objects by visual inspection and processing of many instances/images. The learning process of a human neural model is much more focused on the overall shape and structure rather than focusing on individual features such as a cat's eyes, mouth, ears, and fur. The human brain's core, which is made up of billions of neurons that are wired together, works by abstracting only the necessary data to achieve any task. Hinton et al. \cite{rumelhart_learning_1985} proposed the idea of artificial neural networks that constitutes the foundation of Deep Learning (DL) methods.

Multiple artificial neurons are combined to formalize the hidden layers. These layers are connected to achieve a particular task, as given in Figure  \ref{fig:nn_01}. The intertwining of layers constitutes a dense system of multi-tiered neural network and is central to the concept of deep learning. This method proves to be very useful towards achieving artificial intelligence closer to human learning than the previous ML techniques. DL has many benefits, but it also has some limitations such as it requires a lot of data and needs a lot of computation power to process that data. However, deep learning-based models can be more accurate in numerous fields. These models are now implemented in self-driving cars, weather forecasting systems, and detection of trends in the stock market. However, there is limited use of DL based models in the software design industry. There are open problems within this industry where deep learning algorithms can be applied to automate tasks such as functional software testing, design testing, code generation, and design generation. Here, we discuss fundamental concepts of supervised and self-supervised machine learning strategies. Supervised learning is the method in which labelled data is provided to a training algorithm. On the other hand, in self-supervised learning, the data is provided to the machine without any labels, and the model has to understand the hidden patterns inside the data.

\section{Methods and data to review DL in UI Design synthesis}
\subsection{Study selection using meta-analysis}
A title/abstract/keyword-based search was carried out in the Google Scholar database to find articles related to UI design automation through deep learning techniques. The search query included the “interface design” and “artificial intelligence” (search date: February 23, 2020) keywords, which returned 4,200 results. However, keeping in view the focus of our study on the deep learning-based UI automation systems, we replaced the  “artificial intelligence” with “deep learning” in conjunction with “user interface” and “user interface design”. Consequently, from the query we obtained 374 articles  relevant to our study. 

\subsection{Conferences and Journal Articles}

Table \ref{table:journal_list} and \ref{table:conference_list} show titles and number of relevant papers from journals and conferences related to UI design automation and deep learning. 

\begin{table}[H]
\caption{Journals identified as relevant, and number of reviewed papers.}
\label{table:journal_list}
\begin{tabular}{ll}
\hline
\textbf{Name of Journal}                                                             & \textbf{\#} \\ \hline
ACM Transactions on Graphics                                   & 42                       \\
International Journal of Human-Computer Studies            & 31                     \\
IEEE Transactions on Pattern Analysis and Machine Intelligence & 26                     \\
IEEE Transactions on Software Engineering                      & 11                    \\
International Journal of Human–Computer Interaction            & 7                     \\
ARXIV                                                          & 4                      \\
ACM Transactions on Software Engineering and Methodology       & 2                    \\
Computer Graphics Forum                                        & 3                     \\
Frontiers of Information Technology \& Electronic Engineering  & 3                      \\ \hline
\end{tabular}
\end{table}

\begin{table}[H]
\caption{Conferences/workshops identified as relevant, and number of relevant papers.}
\label{table:conference_list}
\begin{tabular}{ll}
\hline
\textbf{Title of conference/workshop}                                                                                             & \textbf{\#} \\ \hline
Conference on Human Factors in Computing Systems                                                                    & 43                     \\ International Conference on Software Engineering                                                                        & 31 \\
Symposium on User Interface Software and Technology                                                          & 22                     \\
Conference on Pervasive and Ubiquitous Computing & 12 \\
Conference on Computer Vision and Pattern Recognition Workshops & 10 \\
International Conference on Automated Software Engineering & 8 \\
International Conference on Document Analysis and Recognition & 8 \\
\hline               
\end{tabular}
\end{table}

We further reduced the database to 100 articles on the basis of three attributes i.e., “deep learning frameworks”, “precision” and “dataset used”. The resultant database of 100 articles forms the basis of our review presented in Section~\ref{sec_review}.

\section{Background of Deep Learning Methods}
In order to facilitate the reader, we provide a background of deep learning architectures used in UI design automation. For a more detailed overview, we refer to \cite{goodfellow_deep_2016}.\\

\subsection{Building blocks of Deep Networks}
The atomic unit of deep learning models is an artificial neuron that is mathematically represented as a numerical value calculated using a function. 
The parameters to the function are a set of numerical values or matrices, called as weights and biases, denoted by the set $W$ and $B$ respectively.
The parameters are learned by a neural network during training.  
The activations are the non-linear functions, which map the input to a non-linear space and responsible for advancing a neural network from the linear function to a universal function approximator. Sigmoid, Softmax, Tanh, and ReLU \cite{nwankpa_activation_2018} are the most commonly used activation functions in neural networks. 
The optimization process is the backbone of any DL model as it enables the learning of parameters. Gradient descent \cite{ruder_overview_2017} is used as a common optimization algorithm to calculate the gradients of a loss function with respect to parameters and adjust their values iteratively so that the given cost-function is minimized.\cite{rumelhart_learning_1986}. 
The cost function, also known as loss function calculates the error during the training process of the model. Loss functions are of many types, depending on the nature of the task e.g., Mean Squared Error (MSE) \cite{botchkarev_performance_2019} is used for the regression task because it calculates the difference between the observed values (from the dataset) and the predicted values (from the neural network).
\\\\
We will now explore some popular architectures of neural networks commonly used in UI design automation.
\subsection{Deep Neural Networks}
\begin{figure}[]
	\centering
	\includegraphics[width=0.9\textwidth]{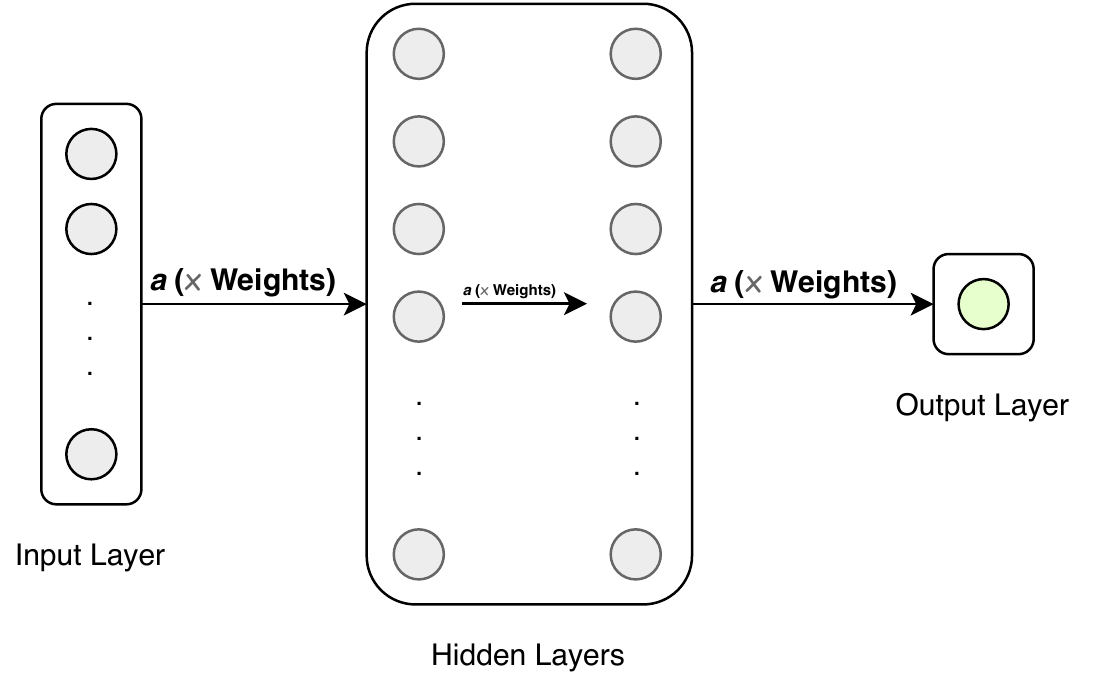}
	\caption{The structure of a deep neural network consists of one input layer, two hidden layers, and a single output layer. Every layer in this network except the final output layer contains n number of nodes. Every node in the layer is multiplied by some weights and passes through an activation function and become the values of the next layer’s nodes.}
	\label{fig:nn_01}
\end{figure}

Deep neural network architectures, inspired from human nervous system design, consist of many layers responsible for information retrieval, concise representation and semantic evaluations. The UI design data provided as input to these dense layers contain different properties of UI design elements, code instructions and images (UI design screenshots). In each layer, a set of linear operations, wrapped by the activation functions $act( )$, the inputs $X$ and parameters $\theta$ are computed. These parameters include weights $W$ and the biases $B$. Equation \ref{eq:nn} shows the formula for an individual neuron ($n_i$).
\begin{equation}
\label{eq:nn}
n_i = act(w_i*x_i + b_i)
\end{equation}
When various neurons are combined in a layer; the values for these neurons are calculated using the weights matrix $W$ and biases $ B $. After the values are computed through hidden layers, they are further connected to the output neurons. These are the predicted outputs against our input data. These values are then compared against observed/target outputs using a proper cost function $c$, such as mean squared error $ \mathbb{E} (Y - \hat{Y}) $, where $Y$ represent labels and $\hat{Y}$ are the predictions. After that, the optimization of the parameters has been performed using the decided optimization process, which is mostly the backpropagation $\delta_L = c' \odot act(W*X + B)'$  \cite{rumelhart_learning_1986}. This process is known as training of a neural network. The aforementioned neural network is also called a fully connected network (FCN) because every neuron in any network layer is connected to all the neurons in its previous layer. A simple structure of the neural network is given in Figure \ref{fig:nn_01}. These neural networks can outperform all other ML models when there is enough data for training. 
\subsection{Convolutional Neural Networks}

The imagery design dataset can be used for classification in various application categories, such as gaming, health-care, and shopping applications. The on-screen element's detection and application content classification can be performed using CNN. Various HCI interfaces utilize user eye movement and gesture detection, whereas convolutional networks can handle dense datasets with a lot of variances.

\begin{figure}[]
	\centering
	\includegraphics[width=0.9\textwidth]{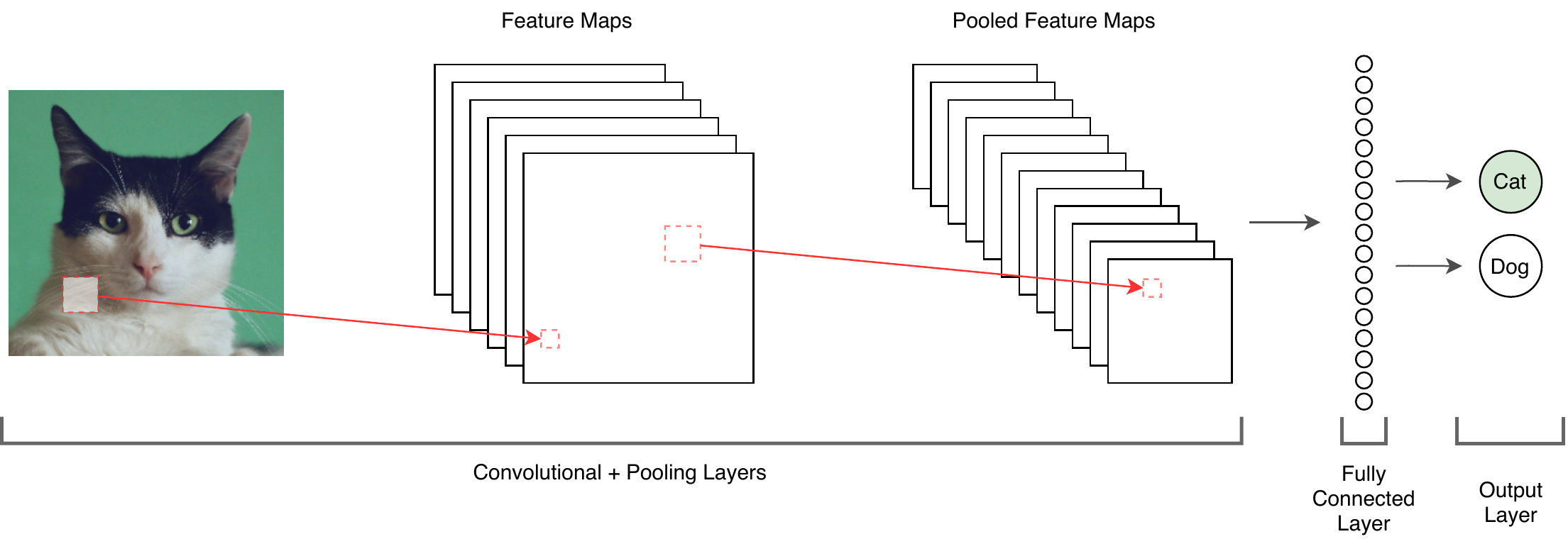}
	\caption{A convolutional neural network having convolved feature maps and pooled feature maps after applying the convolutional layer and pooling. The second last layer of the given CNN is a fully connected layer same as the deep neural network. The CNN in this diagram is performing inference on the picture of a cat. The last output layer contains two nodes which store the probabilities of the input image having a cat or dog.}
	\label{fig:cnn_02}
\end{figure}

The structure and functioning of deep convolutional neural networks (CNN) are inspired by the human visual cortex, which is able to recognize objects by using feature models. In CNN, feature extraction is highly dependent on the filters which are convolved with images to produce feature maps. Consequently, only relevant features are retained in feature maps retrieved after convolutional operations. While working with image data, often stored in the form of multidimensional arrays having different numerical values (0 to 255) at every pixel, the convolution procedure is initiated by selecting parameters such as the size of filters and receptive field are chosen. One may think of the receptive field as a rectangular window of pixels in the image. The data in the receptive field and kernels are convolved together to produce a new single output value of a resultant matrix. Then the receptive field moves further with some value known as a stride. This operation continues until the receptive field completely covers the whole image for all filters. If needed, the padding is also added on the edges of the image so that the information present at the image boundary is also included. This whole operation constitutes one convolutional layer. After that, an optional pooling layer is added in the network to summarize the features in the image, keep the spatial variance of objects in the image, and for reducing the matrix size by choosing either maximum value or mean of values from again a receptive field (in this case, called grid). Subsequently, the data is passed through multiple convolutional and pooling layers. The resultant matrix is then flattened by the network into a vector further connected with the aforementioned DNN. The training process is similar to a deep neural network, noted that in a CNN, the weights are the values in filters, which are learned through the optimization process. A simple architecture of CNN is given in Figure \ref{fig:cnn_02}. Moreover, there are various CNN architectures used for multiple tasks, some popular architectures are LeNet \cite{lecun_gradient-based_1998}, AlexNet \cite{krizhevsky_imagenet_2012}, VGG- 16 \cite{liu_very_2015}, ResNet \cite{he_deep_2016}, GoogLeNet \cite{szegedy_going_2015}.
\subsection{Recurrent Neural Network}
\label{sec:section_rnn}
\begin{figure}[]
	\centering
	\includegraphics[width=0.9\textwidth]{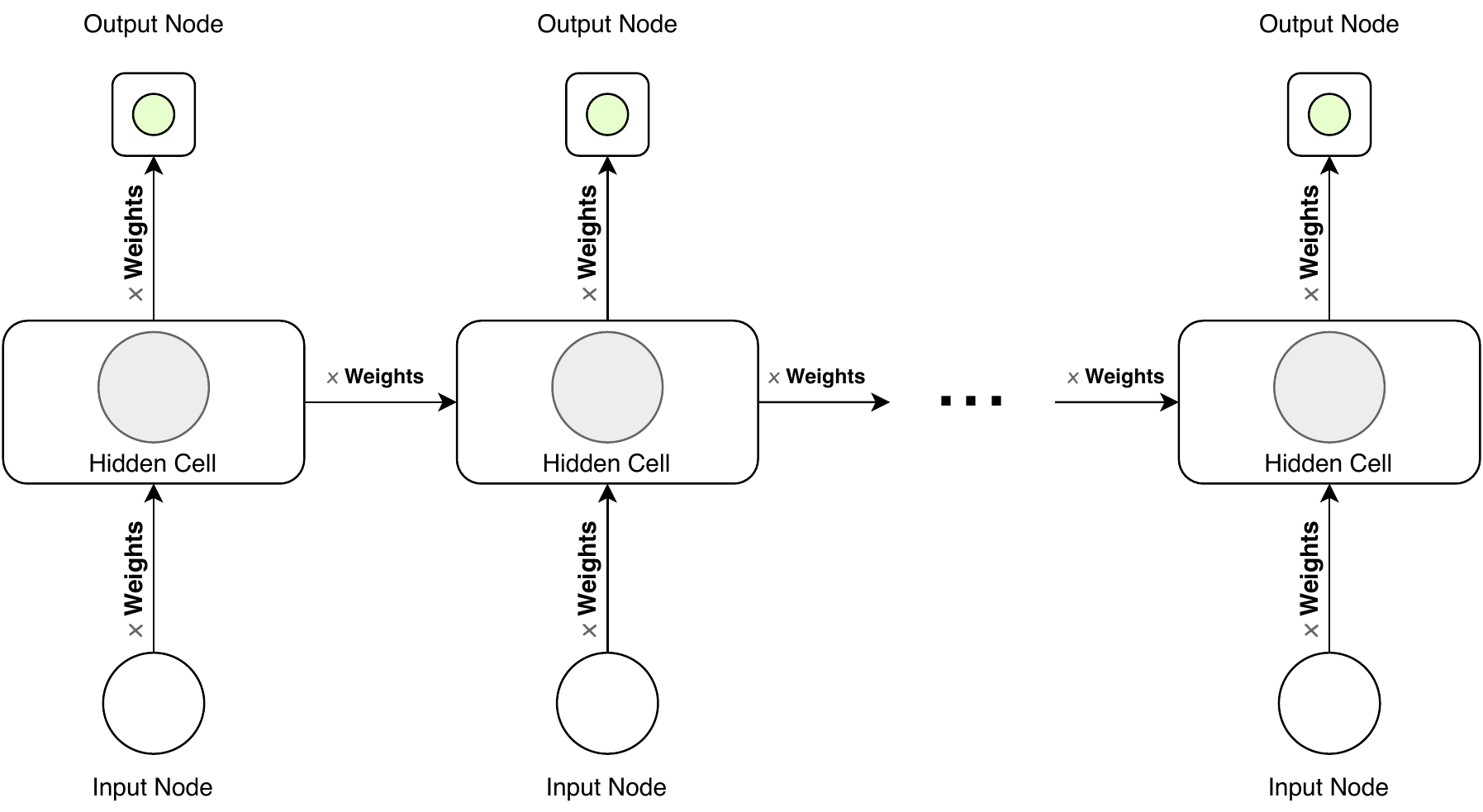}
	\caption{The architecture of a recurrent neural network (RNN) consists of n number of input nodes, but the difference between RNN and DNN is that the hidden nodes (known as hidden cells) are connected. The next node's output is not a pure function of input and activations, but it takes information from the previous “hidden cell” \cite{sherstinsky_fundamentals_2020}.}
	\label{fig:rnn_03}
\end{figure}

Programming language scripts are made up of sequential instructions, and descriptions of UIs contains sequential text data. The sequential data is easily understandable by RNN which learns to query UI design by natural language and is useful for code completion tasks. This can assist software developers and designers while creating new applications.
Recurrent Neural Network (RNN) is a type of sequence model in DL. Sequence models deal with sequential or temporal data. For instance, the reading of English alphabets represents the example of the sequential data. Addressing the alphabets in a series is possible for every speaker, but performing the same task in reverse order may be problematic. The human memory stores the characters in a conditional manner where each next item is dependent on the previous one. RNN follows the same intuition to learn by storing the previous information in a special memory cell and using it with the next state's input for learning and inference. Figure \ref{fig:rnn_03} represents the basic architecture of RNN. In each state of RNN, every neuron is connected with the activation unit (hidden layer) and the activation originating from the previous state. Subsequently, these inputs formulate the output of the current state. In this way, the output of an RNN is not a pure function of inputs, but it is also dependent on previous activations. When the training process of RNN is initialized, the activation unit takes an input from zero vector $a<0>$ considering it a previous unit. Through this structure, RNNs are capable of preserving the information of their internal states. There are two types of weights in RNN, $W_{aa}$ and $W_{ax}$, where $W_{aa}$ is responsible for the transfer of data through the previous activation unit to the current activation unit, and the $W_{ax}$ is for transfer of data from inputs to activation units. RNNs have low computational cost because the weights are shared among all the states. The loss function of a recurrent network is the sum of each loss at every time step. The backpropagation of RNN is named as Backpropagation Through Time \cite{lillicrap_backpropagation_2019}. This backpropagation can mostly cause vanishing and exploding gradient problems. The exploding gradient is when weights become very large and disturb the training of the whole neural network. One way to deal with this problem is the gradient clipping strategy \cite{pascanu_difficulty_2013}. Vanishing gradient problem occurs when the gradients become very small (equal to zero), and training stops. The vanishing gradient issue can be resolved by using various variants of RNN. The most popular and advanced variants of RNNs are Long Short-Term Memory (LSTM) \cite{hochreiter_long_1997} and Gated Recurrent Units (GRU) \cite{cho_learning_2014}.
\subsection{Autoencoders}
\begin{figure}[]
	\centering
	\includegraphics[width=0.9\textwidth]{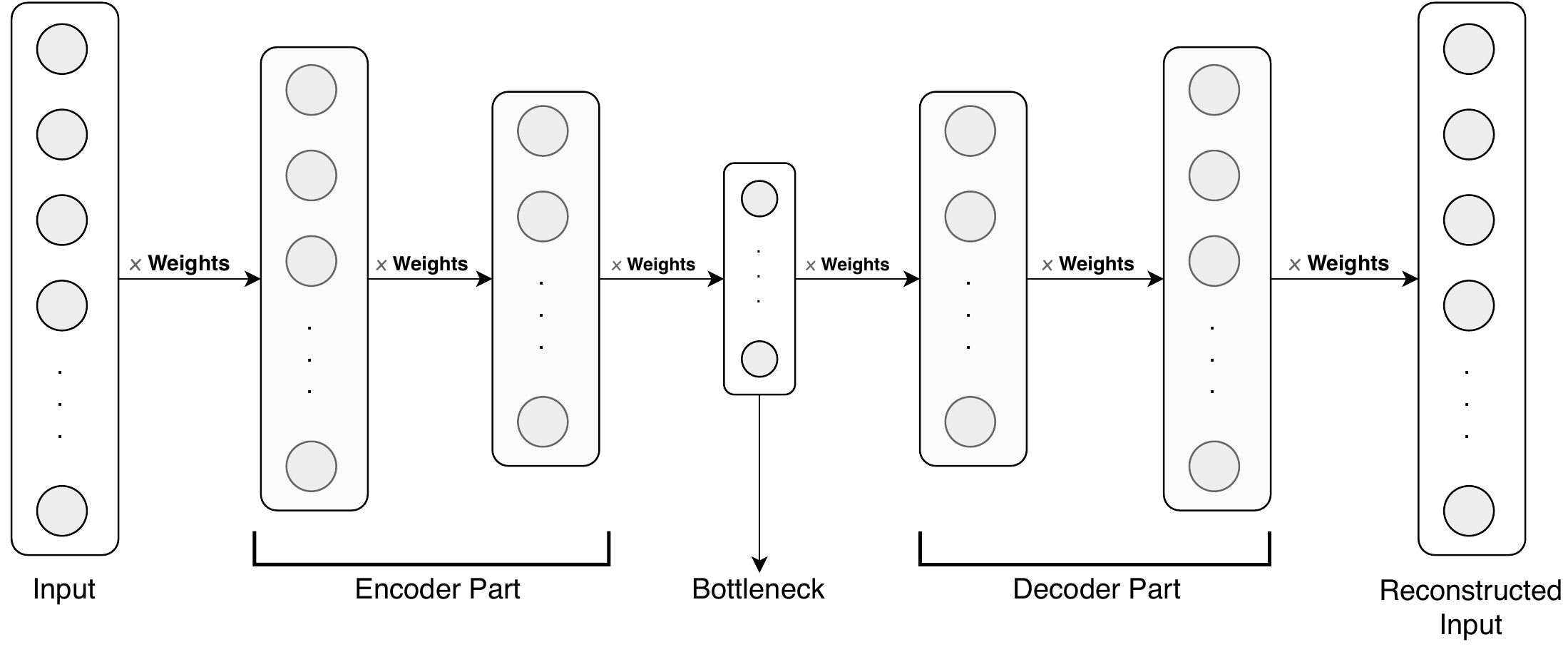}
	\caption{Simple autoencoder model with an encoder and decoder part. Both the encoder and the decoder have two hidden layers. The encoder part receives input and down scales it towards the bottleneck part that contains the compressed representation of the data. This compressed form is then passed through the decoder, which upscales it to reconstruct the input.}
	\label{fig:aed_04}
\end{figure}

Autoencoders belong to the family of self-supervised learning algorithms. Autoencoders encode a compressed representation of data and therefore, used for dimensionality reduction of complex data. In an autoencoder, two different neural networks are joined together via a bottleneck layer (Figure \ref{fig:aed_04}) which is similar to the hidden layer of any neural network. The bottleneck layer is used to compress the input data and plays an important role so that the network should not memorize the input values by mapping them to output neurons. In this way, the number of neurons in the bottleneck layer is much lesser than any other layers in the autoencoder. The left part of an autoencoder ((Figure \ref{fig:aed_04})) is known as the encoder part, which compresses the data into a compact representation with weights, biases, and activation functions. After the data is passed through the encoder and bottleneck layer, it is treated as input for the decoder part for the decompression of the data. An autoencoder's loss function is a reconstruction error, measured by calculating the difference between original inputs and generated outputs (which are recreated inputs). Similar to other neural networks, the backpropagation is used for the training purpose. Autoencoders are comprised of two DNNs, these networks can be any type of neural networks such as convolutional networks as an encoder and convolutional transpose network as a decoder. There are multiple variants of autoencoders for various tasks such as, Denoising Autoencoders \cite{vincent_extracting_2008}, Sparse Autoencoders \cite{jiang_novel_2013}, Stacked Autoencoders \cite{vincent_stacked_2010}, and Variational Autoencoders \cite{kingma_auto-encoding_2014}.
\subsection{Generative Adversarial Networks}
\begin{figure}[]
	\centering
	\includegraphics[width=0.9\textwidth]{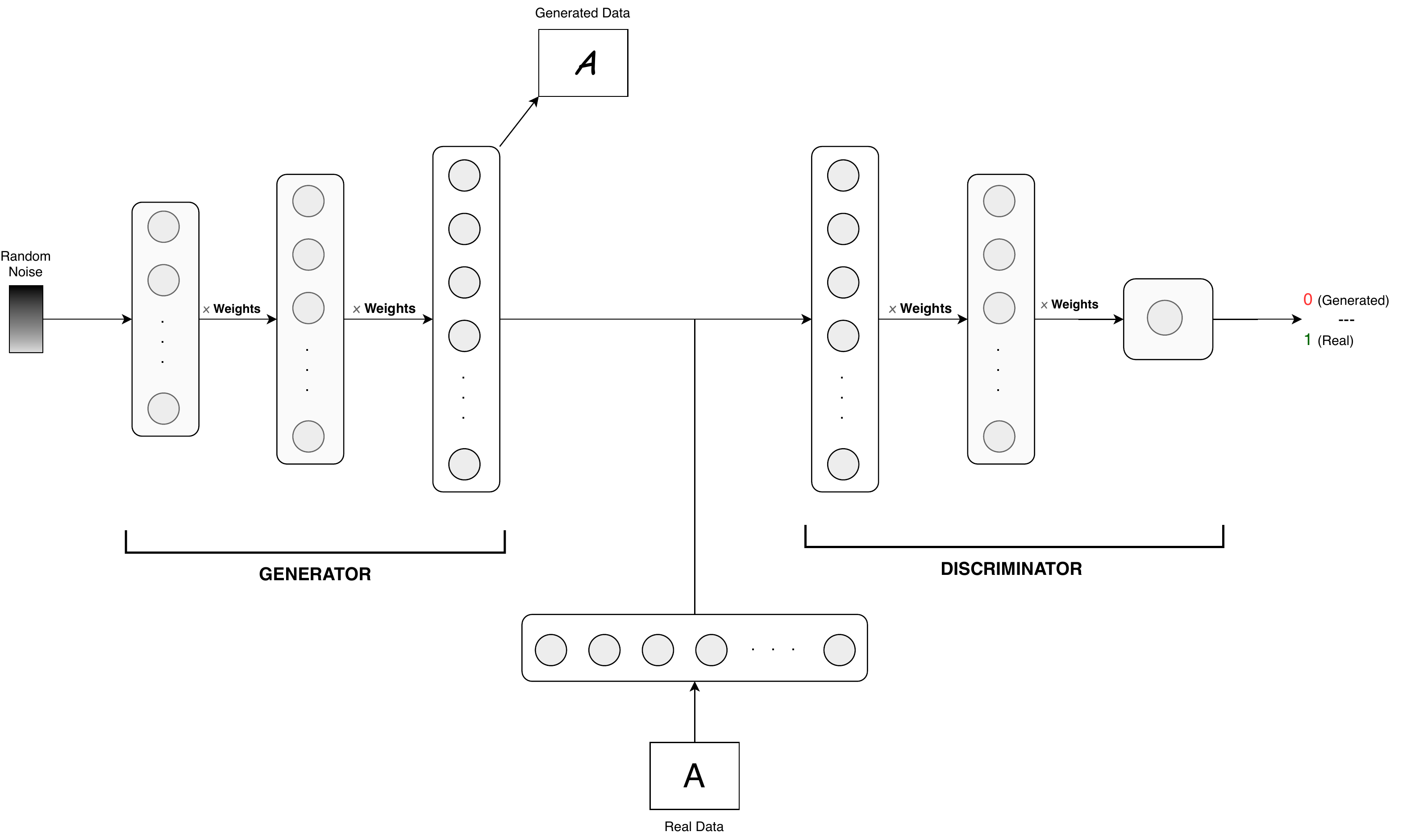}
	\caption{The simple architecture of the generative adversarial network (GAN) in which two neural networks are competing with each other to perform better in the adversary system of minimax game. The generator tries to generate data from random vector space, and the discriminator tries to detect either the data is coming from real data or a generator. In this competition, the generator learns the distribution of real data, and it can generate realistic data.}
	\label{fig:gan_05}
\end{figure}

Generative Adversarial Networks (GANs) are from the family of self-supervised learning algorithms, developed by Ian Goodfellow \cite{goodfellow_generative_2014}. In GANs, two neural networks are competing against each other for their improvement. As displayed in Figure \ref{fig:gan_05}, one is the generator, and the other is the discriminator. The task of the generator is to generate some data given a random noise, and the task of the discriminator is to discriminate between real and generated data. The analogy for understanding the GAN is of considering this network competition between a forger and an investigator. The forger is trying to create fake money, and the investigator has to classify either the money is real or fake. As training continues, both parties keep getting better at their work. When the training finishes, the forger (generator) can generate the money (data) that is indistinguishable from the real money (data). These two networks are competing with each other in an adversarial manner, and this type of learning is called adversarial learning. The cost function of a GAN has two parts; the discriminator's loss function is its ability to distinguish real data from generated data, and the generator's loss function is the disability of the discriminator to recognize generated data. The procedure of training a GAN given as follows: let $\vec{v}$ be the random vector drawn from a Gaussian distribution. The generator takes this $\vec{v}$ as input and tries to map it to the distribution of the real data $R$. During this process, the fake data $F$ is generated by the $G(\vec{v})$. Both the datasets are then classified by the discriminator $D$, such as $D(R) = 1$ and $D(F) = 0$. However, the optimization of the generator's weights is for the matching of fake data to real data, so the generator tries to achieve the $D(F) = 1$. The formulation of loss functions for $G$ and $D$ are given in equation \ref{eq:loss_gan}.
\begin{equation}
\label{eq:loss_gan}
\begin{aligned}
\max L(D;\theta_D) &= \mathbb{E}_{r \sim R} [\log{D(r)}] + \mathbb{E}_{f \sim F}[\log{1 - D(f)}]\\
\max L(G;\theta_G) &= \mathbb{E}_{f \sim F} [\log{D(f)}]
\end{aligned}
\end{equation}
The loss functions represent the maximization process of $ G $ and $ D $; both the neural networks have to optimize their parameters $\theta_G$ and $\theta_D$, respectively. The discriminator's loss function depends on real data, $ R $, and fake data $F$. However, the generator only needs the predictions of discriminator on fake data $D(F)$, where $F = G(\vec{v})$. Using backpropagation, this value is optimized for both networks. The generator has no access to real data, but it still manages to get close towards the actual data, this means that the generator is learning with the feedback from the discriminator. The stability is the biggest challenge of training GANs. If one of these networks outsmarts the other one, the training will fail; for example, if discriminator is overly qualified in discriminating data points, then it will always perform the correct classification, and there is no chance of optimization for generator. Similarly, if the generator becomes over-smart, then it can never achieve diversity in generating data. The ideal point of GAN’s training is to reach an equilibrium point known as Nash equilibrium \cite{nash_equilibrium_1950}.
\section{Datasets for UI design Automation}
\label{sec:dataset}
\begin{figure}[]
	\centering
	\includegraphics[width=0.9\textwidth]{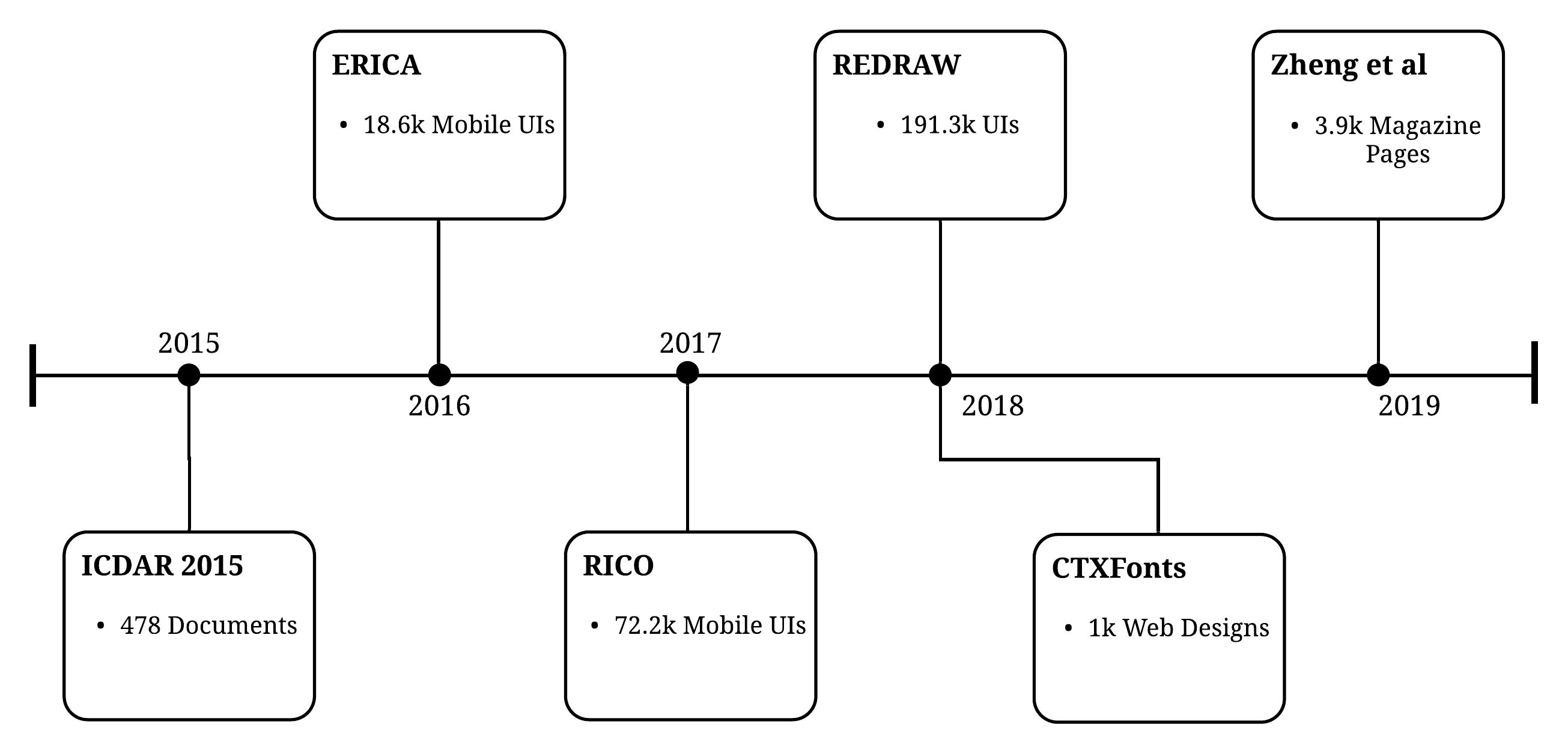}
	\caption{Timeline of user interface related dataset, starting from 2015 to 2019. The figure includes open-source datasets that are useful for deep learning applications. The datasets mostly include the imagery data with some additional features such as labels or text descriptions.}
	\label{fig:dataset_06}
\end{figure}

In this section, we discuss some notable datasets for developing data-driven models for UI automation tasks. In many cases, datasets used by researchers are not publicly available and therefore, we have considered the open-source datasets which are already being used for various applications. Figure \ref{fig:dataset_06} shows the contributions of UI design data sets on a timeline.
\subparagraph{ICDAR 2015}
This is the dataset collected by the Pattern Recognition and Image Analysis (PRImA) \cite{noauthor_prima_nodate} and used in a competition for the segmentation and recognition of textual information on documents. The images of documents contained well-structured layouts of elements such as pictures and text. ICDAR 2015 dataset contains 478 labeled documents \cite{patil_read_2019}. This dataset is used in research for understanding the design of single-page document layouts using DL models \cite{antonacopoulos_icdar2015_2015}.
\subparagraph{ERICA}
This dataset was introduced in 2016 and contained the screenshots of the mobile application’s UIs and user interaction traces on these mobile applications. ERICA has almost 18.6k unique UIs and 50k user interactions collected from 2.4k apps from Google Play Store. ERICA was collected using a dynamic approach when users interact with applications through a web client, and the system captures both the snapshots and user interactions and stores them accordingly \cite{deka_erica_2016}.
\subparagraph{RICO}
Rico is currently the largest repository of mobile application UI designs. It is 4x of the size of ERICA. Rico was presented in 2017 with almost 72.2k snapshots, 10.8k traces of user interactions, and approximately 3 million labeled GUI components from 9.7k applications of Google Play Store. The collection of Rico was in a dynamic manner where users, who are the 13 workers recruited on Upwork, and an automated agent performed mining over the downloaded applications. Rico is currently the most extensive dataset for UIs of mobile applications \cite{deka_rico_2017}.
\subparagraph{REDRAW Dataset}
This is the dataset produced during a study of automating the process of UIs for mobile applications. REDRAW contains 14.3k different screens and 191.3k labeled GUI components. They collected the dataset from the top 250 apps from all the Google Play Store’s categories except games. The GUI elements are labeled in 15 different categories for classification purposes \cite{moran_machine_2018}.
\subparagraph{CTXFonts Dataset}
This dataset was deployed in 2018 and contains screenshots for web designs with labeled text elements. The labels include various font properties such as font face, font size, and font colors. CTXFonts-dataset captures almost 1065 web-designs, 4893 text elements, and 492 font faces, which can be used for various data-driven applications \cite{noauthor_modeling_nodate}.
\subparagraph{Zheng et al. Datasets}
This dataset was open-sourced in 2019 for UI designs and consists of 3,919 magazine page designs. These magazine pages were well aligned with respect to contents. The data set provides semantic annotations of six elements i.e., Text, Image, Headline, Text-over-image, Headline-over-image, and Background \cite{zheng_content-aware_2019}.
\begin{table}
\caption{A table representing the use cases related to datasets, mentioned in section \ref{sec:dataset}. Rico is the most popular dataset, and it was used in multiple UI design studies. Most of the mentioned datasets are used for the empirical validation or testing of the system. ICDAR 2015 is a benchmark dataset for text detection.}
\label{table:ds_use_case}
\begin{adjustbox}{width=\linewidth, center}
\begin{tabular}{ccl}
\hline
Dataset & \multicolumn{2}{c}{Use Case} \\ \hline
\multicolumn{1}{c|}{\multirow{6}{*}{\textbf{RICO \cite{deka_rico_2017}}}} & \multicolumn{1}{c|}{\textbf{Category}} & \multicolumn{1}{c}{\textbf{Description}} \\ \cline{2-3} 
\multicolumn{1}{c|}{} & \multicolumn{1}{c|}{\multirow{2}{*}{Behavioral Studies}} & \begin{tabular}[c]{@{}l@{}}User's response on aesthetics of mobile applications \cite{von_wangenheim_we_2018}.\end{tabular} \\ \cline{3-3} 
\multicolumn{1}{c|}{} & \multicolumn{1}{c|}{} & \begin{tabular}[c]{@{}l@{}}User's concerns about login features in applications \cite{micallef_investigating_2018}.\end{tabular} \\ \cline{2-3} 
\multicolumn{1}{c|}{} & \multicolumn{1}{c|}{\multirow{2}{*}{Empirical Validations}} & \begin{tabular}[c]{@{}l@{}}Object detection on the user interfaces \cite{chen_object_2020}.\end{tabular} \\ \cline{3-3} 
\multicolumn{1}{c|}{} & \multicolumn{1}{c|}{} & \begin{tabular}[c]{@{}l@{}}UI Design querying using autoencoders \cite{chen_wireframe-based_2020}.\end{tabular} \\ \cline{2-3} 
\multicolumn{1}{c|}{} & \multicolumn{1}{c|}{Application Testing} & \begin{tabular}[c]{@{}l@{}}Black-box app testing with deep learning \cite{li_humanoid_2019}.\end{tabular} \\ \hline

\multicolumn{1}{c|}{\multirow{3}{*}{\textbf{\begin{tabular}[c]{@{}c@{}}Magazine \\ Dataset \cite{zheng_content-aware_2019}\end{tabular}}}} & \multicolumn{1}{c|}{\multirow{3}{*}{\begin{tabular}[c]{@{}c@{}}Style-Specific \\ Layout Generation\end{tabular}}} & Single page design generations \cite{zheng_content-aware_2019}.\\ \cline{3-3} 
\multicolumn{1}{c|}{} & \multicolumn{1}{c|}{} & \begin{tabular}[c]{@{}l@{}}Synthesis of the advertising images \cite{you_automatic_2020}.\end{tabular} \\ \cline{3-3} 
\multicolumn{1}{c|}{} & \multicolumn{1}{c|}{} & \begin{tabular}[c]{@{}l@{}}Validation of constraint-based layout generation \cite{lee_neural_2020}.\end{tabular} \\ \hline

\multicolumn{1}{c|}{\multirow{2}{*}{\textbf{\begin{tabular}[c]{@{}c@{}}CTX Fonts \\ Dataset \cite{noauthor_modeling_nodate}\end{tabular}}}} & \multicolumn{1}{c|}{Font Analysis} & \begin{tabular}[c]{@{}l@{}}Comparison of two major font styles, Serif and Sans \cite{shinahara_serif_2019}.\end{tabular} \\ \cline{2-3} 
\multicolumn{1}{c|}{} & \multicolumn{1}{c|}{Font Predictions} & \begin{tabular}[c]{@{}l@{}}Font's properties prediction using deep learning \cite{noauthor_modeling_nodate}.\end{tabular} \\ \hline

\multicolumn{1}{c|}{\multirow{4}{*}{\textbf{\begin{tabular}[c]{@{}c@{}}ReDRAW \\ Dataset \cite{moran_machine_2018} \end{tabular}}}} & \multicolumn{1}{c|}{Code Generation} & \begin{tabular}[c]{@{}l@{}}From UI designs to code generation \cite{moran_machine_2018}.\end{tabular} \\ \cline{2-3} 
\multicolumn{1}{c|}{} & \multicolumn{1}{c|}{\multirow{2}{*}{\begin{tabular}[c]{@{}c@{}}UI Content\\ Analysis\end{tabular}}} & \begin{tabular}[c]{@{}l@{}}Detecting and summarizing of GUI changes \cite{moran_detecting_2018}.\end{tabular} \\ \cline{3-3} 
\multicolumn{1}{c|}{} & \multicolumn{1}{c|}{} & \begin{tabular}[c]{@{}l@{}}Inspires an automated widget recognition system \cite{yu_lirat_2019}.\end{tabular} \\ \cline{2-3} 
\multicolumn{1}{c|}{} & \multicolumn{1}{c|}{Empircial Validations} & \begin{tabular}[c]{@{}l@{}}Used for empirical evaluation of UI object\\ detection system \cite{chen_object_2020}.\end{tabular} \\ \hline

\multicolumn{1}{c|}{\textbf{\begin{tabular}[c]{@{}c@{}}ICDAR \\ 2015 \cite{antonacopoulos_icdar2015_2015}\end{tabular}}} & \multicolumn{1}{c|}{Text Detection} & \begin{tabular}[c]{@{}l@{}}A benchmark dataset for text detection, \\accuracy of 90\% achieved by CharNet \cite{xing_convolutional_2019}.\end{tabular} \\ \hline
\end{tabular}
\end{adjustbox}
\end{table}

\section{Deep Learning in UI Designs}
\label{sec_review}
In this section, we will explore the research already being carried out in the user interface design using DL. Table \ref{table: dataset table} classifies the previous UI automation studies among different deep learning architectures. However, we will sequentially discuss them. We have covered the following attributes for every paper, the DL algorithm used in research, and the dataset used by researchers.\\

Before the development of deep learning, various empirical studies were carried out for the automation of different systems. Biswas et al. \cite{biswas_designing_2012} and Porta et al. \cite{porta_vision-based_2002} reviewed methods and techniques for designing interfaces, while Jalaliniya et al. \cite{jalaliniya_touch-less_2013} built a touch-less interface for interacting with medical images. The automation of designing and the facilitation of easy access to systems other than GUI systems is a promising area of research. Fields in this area include mobile-based augmented reality as well as virtual reality \cite{ye_aranimator_2020, caggianese_freehand-steering_2020, ren_immersive_2010}, text-based insertion \cite{jin_voco_2017}, brain-computer interfaces \cite{zhong_dynamic_2020} and keystroke systems \cite{yang_tapsix_2020}. Aside from this, Liu et al. \cite{liu_understanding_2018} used machine learning algorithms to study the usage patterns of App Store profiles and \cite{zheng_ifeedback_2019, gu_what_2015, wang_e-book_2012, zhang_mallard_2019, ji_uichecker_2018} used them for intelligent system designs and understanding a user's behavior while interacting with UIs. Gollan et al. \cite{gollan_demonstrator_2016} found that user's attention can be maintained by understanding the cognitive load of a person from his pupil dilation. Kolthof et al. \cite{kolthoff_automatic_2019} generated GUI prototypes from text requirements using NLP techniques. Moreover, testing of the android applications was also automated by a framework developed by Wu et al. \cite{wu_exception_2017}.

In 2018 Kevin et al. \cite{moran_machine_2018} develop a tool named REDRAW for the android platform. They implemented various ML techniques in their tool. Their model needs mockup artifacts of designs as an input to their REDRAW pipeline. Firstly, the computer vision techniques are used to detect the GUI components in screenshots of multiple android apps; after that, CNNs were implemented to classify the identified components accordingly. Finally, they used the k-nearest Neighbor (KNN) to assemble the elements precisely. The dataset in this work was built on the top 250 Android applications and collected about 14,382 different screens and 191,300 labeled interface components. Nguyen et al. \cite{nguyen_deep_2018} in 2018, developed a system for UI design automation. Firstly, a RNN is trained to query UIs based on text descriptions. They used these gathered UIs to generate UI designs using GANs. The authors had created the dataset of multiple applications UIs and manually added the descriptions of these UIs. Kevin et al. proposed two different models: (i) NaturalUI, which can query design using textual query; (ii) GenUI generates images of UIs. In 2018 Liu et al. \cite{liu_learning_2018} used various DL techniques to create a model that could assist UI designers. The dataset of labeled UI components had been created, and then trained a CNN on that dataset to differentiate the icon classes. During this study, an autoencoder was trained to learn the vector representation of the Rico dataset's semantic layouts, which was later used with the KNN to query UIs. However, the decoder part of the trained autoencoder can generate UIs based on its learning. This work is only for the user interface designs of android applications. Zhao et al. \cite{noauthor_modeling_nodate}, in 2018, developed a multitask neural network to predict the font properties of text elements in web interfaces. Font properties include three things font face, font color, and font size. They worked on a dataset of 1k web designs, having labeled text elements against their font properties. The CNN and autoencoders were used along with the adversarial training process to create a novel predictive model.\\ Wu et al. \cite{wu_understanding_2019} proposed a study to perceive mobile application rating using a machine learning-based model. 318 mobile apps from the Rico dataset were selected for this study. The freelancers were hired to rate these applications based on colors, textures, and layouts; then they produced a predictive model based on the labeled dataset created by the freelancers. They used 5 machine learning models, while carrying out a comparative study on these models for the design rated perceptive. They used the following models: i) multiple linear regression ii) lasso linear regression iii) multi-layer perceptron iv) decision tree v) random forest for the prediction of the personality of apps. Their analysis is based on 15 different variables and five dimensions, which can affect any device's ranking. In May 2019, Huang et al. \cite{huang_swire_2019} used several ML methods to develop a system that can retrieve the UIs of mobile application using the sketches as input. These ML techniques include the Bag of Words (BOW) and the Histogram of Oriented Gradients (HOG) filters. Their contribution is the detection of similarities between the original UI and the sketch image. The developed product was a UI querying system that was built by the combination of DNN and KNN to scan for similar UIs, rank them accordingly, and display them in order. The querying framework cannot work with the complex UI systems. Rico dataset was used to select the interfaces and create the sketches of these interfaces. In 2019 Zheng et al. \cite{zheng_content-aware_2019} used the strategy of GANs to generate magazine layouts. The generator works together with three different encoders for accurate generation or alignment of elements in the interfaces. They picked three elements, such as pictures, keywords, and type (category) of the document. These three elements have their encoders, which operate with the GAN's generator to trick the discriminator. The model learned the structure and elements from layouts. This research was not specific for mobile applications, as the dataset of magazine documents had been created by using masks produced by semantic segmentation of the element. Patil et al. \cite{patil_read_2019} generated UI designs with a combination of the Variational Autoencoders (VAEs) and the RNNs. The dataset used in this study is of single-page documents named ICDAR explained in the datasets section \ref{sec:dataset}. A new metric was introduced for measuring the diversity and uniqueness of UI layouts. The RNN is used to breaks down the relative element in layouts into various chunks. These elements are then treated as sequential data for the training of a variational autoencoder (VAE), comprised of Spatial-relation Encoder and Decoder (SRE / SRD). However, this system was unable to handle the complex structures of UI designs. Gaussian distribution was used as the weights initializer to mitigate the problem of vanishing and exploding gradients. Li et al. \cite{li_layoutgan_2018} utilized the GANs to generate the layouts of mobile applicaions by understanding how their layouts were organized to the graphic elements in them. Using DCGAN, the models learn from semantic data masks; they created two distinct discriminators that are based on two separate techniques: (i) relation based (ii) wireframe. In relation-based discriminator, the model tries to understand the relationship between the elements using different classifiers. In the latter, the wireframes are rendered by converting the images of the dataset to grayscale, and then, these wireframes are trained on a CNN to discriminate the real/fake images. 25k templates were built from existing documents and validity was checked on Rico semantic layout wireframes. Yet one thing to note here is that this work is not limited to smartphone devices or webpages. Lee et al. \cite{lee_design_2019} used the Conditional GANs (CGANs) to generate the color palette for the interfaces of the Android applications, using the design semantics as conditions for CGANs. The analysis, in this study, yields promising results, but the research covers just a single component of the UI, color. The authors had gathered the GUI design dataset from LG Electronics to train their network. Schlatter et al. \cite{schlattner_learning_2019, chen_fromui_2018} developed a method for pixel-accurate implementation of any UI design from its screenshot. They cover various UI elements e.g., border properties, frame properties, color properties, and text properties. Imitation learning achieved up to 94.8\% accuracy in inferring the value of attributes in the given image. A new cost function had been proposed for estimating the values of attributes. The methodology was: First, the picture goes to model and passes through several CNNs, then each CNN detects the quantity of the attributes in the images. These values are used to regenerate/render the interface after CNN's predictions; these generated layouts are then compared to the original ones for training. A dataset was collected from Google Play Store's Android applications, and a synthetic dataset was also produced while training. This work was a positive step towards the design automation work.\\ Lee et al. \cite{lee_guicomp_2020} developed a tool called GUI-Comp, which works as an extension of the KakaoOven tool to assist graphic designers with tips and feedback in real-time. This program had been tested against inexperienced designers. GUI-Comp includes three panels to support the designers. Recommendation panel: trained on the Rico dataset, utilizing the stacked VAEs along with KNNs to recommend the elements that should be used for interface architecture. Attention Panel: Uses a pre-trained, fully connected neural network (FCN) to predict the user's attention on the created design. The FCN was trained on graphic design importance (GDI) dataset for providing the heatmap that reflects the user's attention. Evaluation Panel: it contains multiple metrics that compute the overall score of design based on the element's properties and adjustments.
\begin{table}
	\renewcommand\arraystretch{1.0}
	\newcolumntype{C}{ >{\centering\arraybackslash} m{4cm} }
	\newcolumntype{D}{ >{\centering\arraybackslash} m{1cm} }
	\centering
	\caption{Studies classified by the deep learning methods }
    \begin{adjustbox}{height=3.4in}
	\begin{tabular}{|c|l|l|}
		\hline
		\multicolumn{3}{|c|}{\textbf{Supervised Learning}}                                                                                                                                          \\[2ex] \hline 
		\multirow{18}{*}{\textbf{\begin{tabular}[c]{@{}c@{}}Deep \\ Neural Network \\ (DNN)\end{tabular}}}                                   & Huang et al. \cite{huang_swire_2019}        & Sketch-based querying of UIs                                        \\ \cline{2-3} 
		& Lee et   al. \cite{lee_guicomp_2020}        & Predict the user’s attention on a   UI                              \\ 
		\cline{2-3} & Villegas et al. \cite{villegas_neural_1994}       & Identify cognitive goals of users      \\
		\cline{2-3} & Kapoor et al. \cite{kapoor_automatic_2007}        &   User's interaction frustation prediction   \\
		\cline{2-3} & Vizer et al. \cite{vizer_automated_2009}        & Stress detection using keystrokes      \\
		\cline{2-3} & Von et al. \cite{van_tonder_improving_2012}        & controlled interaction for map-based applications      \\
		\cline{2-3} & Lian et al. \cite{lian_easyfont_2018}        & Cutomized font creation      \\
		\cline{2-3} & Li et al. \cite{li_applications_2017}        & review on intelligent manufacturing      \\
		\cline{2-3} & Liu et al. \cite{liu_deep_2020}        & Study of AI-based software generation      \\
		\cline{2-3} & Ma et al. \cite{ma_easy--deploy_2019}        & API extraction using transfer learning      \\
		\cline{2-3} 
		& Wu et al. \cite{wu_understanding_2019}         & Prediction of app score                                             \\
		\cline{2-3}& Wenyin et al. \cite{liu_wenyin_smart_2001} & Shapes recognition from sketches \\
		\cline{2-3}& Zou et al. \cite{zou_overlapped_2011} & Framework for overlap hand-writing in smartphone UIs.  \\
		\cline{2-3}& Ahmed et al. \cite{ahmed_targeted_2019} & Error-based code generation \\
		\cline{2-3}& Lavania et al. \cite{lavania_weakly_2016} & Biological laboratory assistance using DNN \\
		\cline{2-3}& Wen et al. \cite{wen_ubitouch_2016} & Smart touchpads for smartphones \\
		\cline{2-3}& Jungwirth et al. \cite{jungwirth_mobeyele_2019} & Industry's worker assistance by eye tracking \\
		\cline{2-3}& Kim et al. \cite{kim_say_2019} & Smart wearable device for visually impared people \\ 
		[2ex]\hline
		\multirow{16}{*}{\textbf{\begin{tabular}[c]{@{}c@{}}Convolutional \\ Neural Network \\ (CNN)\end{tabular}}}                           & Moran et al. \cite{moran_machine_2018}        & UI element’s assembling after   detection                           \\
		\cline{2-3} & Dou et al. \cite{dou_webthetics_2019}       & Quantification of website aesthetics    \\
		\cline{2-3} & Halter et al. \cite{halter_vian_2019}       & Annotation tool for films    \\
		\cline{2-3} & Bao et al. \cite{bao_psc2code_2020}       & Programming code extraction    \\
		\cline{2-3} & Han et al. \cite{han_deepsketch2face_2017}       & 3D sketching system    \\
		\cline{2-3} & Bell et al. \cite{bell_learning_2015}       & product visual similarity    \\
		\cline{2-3} & Nishida et al. \cite{nishida_interactive_2016}       & sketching of urban models    \\
		\cline{2-3} & Shao et al. \cite{shao_interactive_2012}       & Semantic modeling of indoor scenes    \\
		\cline{2-3} 
		& Schlattner et al. \cite{schlattner_learning_2019} & Prediction of the element’s   property value                        \\
		\cline{2-3} & Bylinskii et al. \cite{bylinskii_learning_2017} & User's focus areas prediction \\
		\cline{2-3} & Liu et al. \cite{liu_learning_2018} & Created semantic annotations for Rico dataset. \\
		\cline{2-3} & Yeo et al. \cite{yeo_opisthenar_2019} & Pose recognition by using wearable device \\
		\cline{2-3} & Kong et al. \cite{kong_selecting_2016} & Smart glass UI for the selection of home appliances \\
		\cline{2-3} & Mairittha et al. \cite{mairittha_improving_2020} & Mobile UIs personalization detection and perdiction \\
		\cline{2-3} & Stiehl et al. \cite{stiehl_towards_2015} & UI for sign wirting (hand gesture) detection \\
		\cline{2-3} & Tensmeyer et al. \cite{tensmeyer_convolutional_2017} & Font recognition and classification \\
		[2ex] \hline
		\multirow{9}{*}{\textbf{\begin{tabular}[c]{@{}c@{}}Recurrent \\ Neural Network \\ (RNN)\end{tabular}}}
		 & Wang et al. \cite{wang_vision-language_2020}       & Navigation using natural language \\                            
		 \cline{2-3} & Fowkes et al. \cite{fowkes_autofolding_2017}       & Automatic folding of IDE code \\ \cline{2-3}                  & Nguyen et al. \cite{nguyen_deep_2018}       & Query UI with textual input                                         \\
		\cline{2-3} & Mohian et al. \cite{mohian_doodle2app_2020} & Code generation from UI sketches \\
		\cline{2-3} & Zhang et al. \cite{zhang_type_2019} & Automated smart text correction system \\
		\cline{2-3} & Huang et al. \cite{huang_sketchforme_2019} & Sketch generation from natural language descriptions   \\
		\cline{2-3} & Sun et al. \cite{sun_lip-interact_2018} & Controlling mobile functions from lip movement \\
		\cline{2-3} & Zhou et al. \cite{zhou_lancer_2019} & Code completion based on previous code \\
		\cline{2-3} & Bhatt et al. \cite{bhatt_digital_2019} & Invoice document structure recognition \\
		 [2ex] \hline
		\multicolumn{3}{|c|}{\textbf{Self-Supervised Learning}}                                                                                                                                     \\[2ex] \hline
		\multirow{8}{*}{\textbf{Autoencoders}}                                                 & Liu et al. \cite{liu_learning_2018}          & Understand the representation of   UIs in Rico dataset              \\ \cline{2-3} 
		& Patil et al. \cite{patil_read_2019}      & Variational Autoencoders to   generate single page UI documents     \\\cline{2-3} 
		& Tufano et al. \cite{tufano_empirical_2018}      & Bug fixing using neural machine translation     \\
		\cline{2-3} & Lekchas et al. \cite{lekschas_peax_2020}      & Unsupervised technique for visual pattern search  \\ \cline{2-3} 
		& Lee et al. \cite{lee_guicomp_2020}        & Stacked VAEs to recommend the   elements while creating a UI design \\ \cline{2-3} 
		& Zhao et   al. \cite{noauthor_modeling_nodate}       & Font Prediction on web designs   with adversarial training          \\
		\cline{2-3} & Chen et al. \cite{chen_fromui_2018} & Automated GUI reconstruction from pre-designed UI's image \\
		\cline{2-3} & Ge et al. \cite{ge_android_2019} & Query android UIs from sketches \\
		[2ex] \hline
		\multicolumn{1}{|l|}{\multirow{4}{*}{\textbf{\begin{tabular}[c]{@{}c@{}}Generative \\ Adversarial Network \\ (GAN)\end{tabular}}}} & Nguyen et al. \cite{nguyen_deep_2018}       & GenUI to generate UI designs from   the custom dataset              \\ \cline{2-3} & Zheng et   al. \cite{zheng_content-aware_2019}      & Generation of magazines by given   images, text and category    \\ \cline{2-3}             & Li et al. \cite{li_layoutgan_2018}         & Generate layouts of UI designs   using various Discriminators       \\
		 \cline{2-3} & Zhang et al. \cite{zhang_two-stage_2018} & Software for colorization of arts    \\
		 \cline{2-3} & Sun et al. \cite{sun_smartpaint_2019} & Automated drawing system trained on cartoon images     \\
		 \cline{2-3} & Zhang et al. \cite{zhang_anticipating_2019} & People attention detection using advesarial networks    \\
		 \cline{2-3} & Lee et al. \cite{lee_smartphone_2019} & Conditional GANs for mobile content organization    \\
		 \cline{2-3} 
		& Lee et al. \cite{lee_design_2019}        & Conditional GANs to generate the   color palette for apps           \\[2ex] \hline
	\end{tabular}
	\label{table: dataset table}
	\end{adjustbox}
\end{table}

\section{Discussion and Research Frontiers}
\label{sec:discussion_research_frontiers}
In this paper, we reviewed important contributions in the area of DL towards UI design. We discussed DL architectures across emerging methods ranging from the DNNs to GANs. Since each previous contribution involves different datasets, preprocessing methods, performance metrics, and DL models, it is challenging to generalize and draw conclusions about the performance of any particular approach. Therefore, the comparison presented in this paper encompasses different methods and their reported performance on their given dataset.\\
The CNNs based element- wise classification has reached almost to the 95\% accuracy \cite{moran_machine_2018, schlattner_learning_2019}. Similar to the CNN based models, the querying approaches that use RNNs have comparable performance with an additional benefit that they can adapt to textual data as well as design sketches. Unlike the discriminative studies, the contributions that use generative modeling \cite{patil_read_2019, noauthor_modeling_nodate, zheng_content-aware_2019, nguyen_deep_2018,  liu_learning_2018, li_layoutgan_2018, lee_design_2019, lee_guicomp_2020} for UI generation tasks are also giving promising results.\\

However, there are still many gaps in the design automation area of human-computer interaction. The utilization of DL frameworks can be used as a solution for various open research problems. The detailed discussion of these problems with the possible solutions and essential future directions in this area are discussed below.

\subsection{Repositories of Cross-Platform Designs}

Large and frequently revised benchmark datasets are critical to advance the design automation process to gain real trust in DL methods. A high-quality imagery dataset is available for only android mobile applications \cite{deka_rico_2017}. However, the area of cross-platform interface designs is also uncovered. The research for iOS and desktop application designs is very limited. The website designs are also ignored to an extent. This gap can be bridged by the iOS user community to build a dataset or make the already existing design dataset of interfaces available, so researchers can contribute and explore this area on the iOS platform. There is also a need for cross-platform UI/UX dataset. The cross-platform designs are necessary because Android OS has only a 36\% market share. The other operating systems are also important and can not be ignored; from the statistics of \cite{noauthor_operating_nodate}, it can be seen that Windows, iOS has 36\% and 14\% market shares, respectively. Data-driven applications can only be possible by using these datasets. The huge repositories are very much needed for the support of DL models. 

 \subsection{Advanced UI Generation}
Current research in UI design generation is very limited. With the availability of generative models, the data generation process is automated in every industry. A team of scientists and researchers, inspired by MIT, work specifically on the use of generative modeling in different fields of science and arts \cite{noauthor_how_nodate}. They have been working for the generation of perfumes, viruses, music, and fashion. These advance usages of generative modeling can also be used for the generation of advanced UI designs. The dynamic elements can also be treated as variables in the design process because the recent work carried out in GANs can generate high definition images with control over features or elements. Apart from this High-quality UI content generation, the requirement specification process can also be changed using the most recent contributions in the field text to image translation such as \cite{qiao_mirrorgan_2019}. The research should not be limited to only image generation methods, but it can be advanced to generate the code for various design languages. These codes can then be modified by various NLP architectures, described in section \ref{sec:section_rnn}, using human-friendly textual input.

\subsection{Centralized System}
There is also a need for the development of a unified standard in the form of an open-source, cross-platform DL UI design automation framework, which could have various DL strategies implemented in it while covering multiple areas of UI systems. This envisaged framework, if developed, will have both elementwise and layout-wise implementation of DL models that allow users to customize the designs as needed. This system may be in the form of a web application or desktop application. However, it should have the ability to work with the pre-build UI/UX systems such as Adobe XD \cite{noauthor_uiux_nodate}, InVision \cite{noauthor_invision_nodate} and Sketch \cite{noauthor_digital_nodate}. The idea of the centralized UI system can also be extended for the user experience (UX) research using various ML and DL methods.\\ 

The unawareness of novice designers is a reason for the limited use of DL-based UI design automation methods. This problem can be minimized if the popular software development tools such as Android Studio \cite{noauthor_meet_nodate}, XCode \cite{noauthor_xcode_nodate}, and Microsoft Visual Studio \cite{noauthor_visual_nodate} allow the integration of DL based designing models in them.

\section{Conclusions and Future Works}
In this paper, we presented a comprehensive review of the existing work using deep learning methods towards UI design. Our study involves deep neural networks, convolutional neural networks, recurrent neural networks, autoencoders, and generative adversarial networks. We reviewed different data sets that have been used along with the aforementioned deep learning techniques. Some successful use cases related to UI datasets are discussed in this study. We also discussed some other interface designs that may somehow be related to software designs, such as one-page magazine layouts that are similar to mobile app UI designs. 
In section \ref{sec:discussion_research_frontiers}, we outline some future application areas of UI design automation by using deep learning algorithms. Some important futures extensions of the existing work may include: (i) automation of software testing and user experience, (ii) collection of cross-platform UI design datasets, (iii) application of semi-supervised generative algorithms for synthesis of UI components, (iv) improvement in explainability and interpretability of UI content generation approaches.  The use of deep learning for UI design automation tasks could be one of the high potential fields for the advancement in software development, but it is still unexplored. Based on the current and foreseeable future applications, we believe that the deep learning models will transform the software industry's landscape in general and application design in particular.

\bibliographystyle{unsrtnat}
\bibliography{paper.bib}
\end{document}